\documentclass{PoS}

\usepackage{graphicx}

\newcommand{\new}{\newcommand}
\new{\CERN}        {\mbox{\small\textsc{CERN}}}
\new{\LHC}        {\mbox{\small\textsc{LHC}}}
\new{\SLHC}        {\mbox{\small\textsc{SLHC}}}
\new{\ATLAS}        {\mbox{\small\textsc{ATLAS}}}
\new{\CMS}        {\mbox{\small\textsc{CMS}}}
\new{\TOTEM}        {\mbox{\small\textsc{TOTEM}}}
\new{\ECAL}        {\mbox{\small\textsc{ECAL}}}
\new{\LHCb}        {\mbox{\small\textsc{LHC}b}}
\new{\ALICE}        {\mbox{\small\textsc{ALICE}}}
\new{\LEP}        {\mbox{\small\textsc{LEP}}}
\new{\DELPHI}        {\mbox{\small\textsc{DELPHI}}}
\new{\Lthree}        {\mbox{\small\textsc{L3}}}
\new{\TEVATRON}        {\mbox{\small\textsc{TEVATRON}}}
\new{\FERMILAB}        {\mbox{\small\textsc{FERMILAB}}}
\new{\HERA}        {\mbox{\small\textsc{HERA}}}
\new{\RHIC}        {\mbox{\small\textsc{RHIC}}}

\title{LHC Expectations (Machine, Detectors and Physics)}

\ShortTitle{LHC Expectations (Machine, Detectors and Physics)}

\author{\speaker{G\"unther Dissertori}
              \\
        Department of Physics, ETH Zurich, Switzerland\\
        E-mail: \email{guenther.dissertori@cern.ch}}

\abstract{
Starting in two years from now, particle physics will enter a new regime
in terms of energies and luminosities, thanks to the Large Hadron Collider (\LHC) at CERN.
This report summarizes the status of the preparations, both for the machine
and the detectors, as of fall 2005. The commissioning and start-up scenarios are outlined and
some highlights from the very rich physics programme are given, concentrating on measurements of
Standard Model processes, as well as on early discovery scenarios. The prospects
of B-physics and heavy ion collisions at LHC are also briefly discussed.
The report concludes with an outlook on the ultimate physics reach and on upgrade scenarios.
}

\FullConference{International Europhysics Conference on High Energy Physics\\
		 July 21st - 27th 2005\\
		 Lisboa, Portugal}

\begin{document}

%%%%%%%%%%%%%%%%%%%%%%%%%%%%%%%%%%%%%%%%

\section{Introduction}
 \label{intro}
 
 We  are approaching the start-up of the world's most powerful particle accelerator
 ever built. In about two years from now, \CERN's Large Hadron Collider (\LHC) \cite{LHC} will
 starts its operation. Thanks to the unprecedented energies and luminosities, it
 will give particle physicists the possibility to explore the TeV energy
 range for the first time and hopefully discover new phenomena, which go beyond the
 so successful Standard Model (SM). 
 
 The physics motivations for this new endeavour are manifold. Above all, it is
 believed that the origin of electro-weak symmetry breaking  will be elucidated. 
 Concretely speaking, this might consist in the discovery of one or more Higgs
 bosons and thus confirm the prediction that there is spontaneous symmetry 
 breaking via the Higgs mechanism \cite{Higgs}. On the other hand, if no evidence 
 is obtained for a Higgs mechanism, we nevertheless expect new phenomena to show
 up in the TeV energy range,
 which after all have to ensure the conservation of unitarity. The latter is
 known to be violated, for example in the scattering of the longitudinal components
 of two W bosons, if no new phenomena set in at the TeV scale. 
 
 The other main field of activity will be the search for new types of symmetries
 and particles, most notably Supersymmetry (SUSY). We refer to \cite{SUSY} for an overview
 of SUSY and its phenomenological implications. SUSY  is the most prominent and
 carefully studied model of all proposed extensions of the SM. It postulates a
 symmetry between fermions and bosons and introduces a rich new spectrum of
 particles. This theory has several strong motivations.
 For example, it proposes a rather natural solution of the hierarchy problem, 
 if supersymmetric partners of the SM particles appear with masses below or around the
 TeV scale. This would prevent the Higgs mass to acquire enormously large radiative corrections
 and eliminate the need to have an unnatural fine tuning in order to explain the apparently
 small Higgs mass. The appearance of SUSY particles would also lead to the convergence
 of the electro-weak and strong coupling constants at an energy of about $10^{16}$ GeV,
 as generally foreseen in scenarios of Grand Unified Theories (GUT). Finally, some implementations
 of SUSY provide an excellent candidate for the dark matter observed in our Universe, namely
 the weakly interacting and stable lightest neutralino.
 
 Recently other solutions for the hierarchy problem have been put forward, which postulate
 the existence of Extra Dimensions (ED). Some of these models \cite{ADD} conjecture that
 the fundamental scale of gravity could be as low as the TeV scale. The Planck scale
 only appears as a derived scale  because of the large volume of the EDs and the
 fact that only gravity can propagate there, whereas all SM fields are confined to a
 four-dimensional brane. Thus we see only a small part of the total gravitational flux, which
 explains the relative weakness of gravity compared to the other SM interactions. Other
 models exist \cite{RS} which try to explain the hierarchy problem by a very strong curvature of
 the EDs. In general, the phenomenology of EDs foresees the production of gravitons via
 parton scattering at the \LHC. These gravitons could escape into the EDs, leading to an 
 apparent violation of energy-momentum conservation in our four-dimensional world, or
 decay to SM particles in a resonant-like behaviour. 
 
 Of course, at \LHC\ we will also search for new interactions and their related carrier particles,
 such as new vector bosons (Z', W') with masses of a few TeV/$c^2$. These particles
 arise in models which extend the gauge group of the SM, as for example
 the recent Little Higgs models (see eg.\ \cite{LittleHiggs} and references therein). 
 Another proposal to explain electro-weak symmetry
 breaking is given by Technicolor models \cite{Technicolor}. Also there we would expect
 the appearance of heavy resonances. 
   
 Besides the direct searches for physics beyond the SM, precision studies of the heavy flavour
 sector could lead to indirect evidence for new physics, for example via an enhancement
 of otherwise very rare decays. In particular, the very copious
 production of $\mathrm{B}_\mathrm{s}$ mesons will allow to complement the 
 measurements at the B-factories currently in operation. The flavour mixing
 parameters which appear in the Cabbibo-Kobayashi-Maskawa (CKM) matrix will
 be measured using several different decay channels, hopefully leading to a better understanding
 of CP-violation in the B-sector.
  
 Finally, the \LHC\ will also allow to collide heavy ions. The unprecedented energy densities
 achieved in these collisions are expected to lead to the formation of new forms of
 partonic matter, most notably a quark-gluon plasma. The properties of this new state of 
 matter, as well as the phase transition to hadronic matter will be the subject of an intense
 research.
 
 This very rich physics programme will be pursued at \CERN\ by observing proton-proton
 collisions (as well as heavy ion collisions) at four experimental sites around the
 \LHC\ ring. It will be installed in the former \LEP\ tunnel, about 100 m underground with a
 circumference of approximately 27 km. Two of the four experiments, \ATLAS\ and \CMS,
 will be large general purpose detectors designed to cover practically the whole
 range of physics questions outlined above. The other two experiments 
are optimized for the study of B-physics (\LHCb) and heavy ion collisions (\ALICE).

In the following we will summarize the status of the preparations of the machine and the
detectors and describe the planned start-up scenarios. In a more detailed discussion
of the foreseen physics analyses, emphasis will be given to the early physics reach.
The report is concluded with an outlook on the later physics reach and possible \LHC\
 upgrade scenarios. Other reviews of physics at the LHC and its preparations can be found in Refs.\ 
 \cite{FabiolaHinchcliffe}, \cite{FabiolaLP05} and \cite{GigiLP05}.

%%%%%%%%%%%%%%%%%%%%%%%%%%%%%%%%%%%%%%%%

\section{Construction status}
 \label{status}

%%%%%%%%%%%%%%%%%%%%%%%%%%%%%%%%%%%%%%%%

\subsection{Status of the LHC construction}
 \label{status-LHC}

The LHC will be a proton-proton collider with an energy per beam of 7 TeV, a factor of
seven larger than the currently highest energy achieved in the world, namely with the
\TEVATRON\ at \FERMILAB. Its main components will be 1232 superconducting dipoles,
each 14.2 m long (magnetic length).
When operated at their nominal temperature of 1.9 K, a magnetic field of 8.33 T is achieved.
They are of the "2 in 1" type, meaning that the apertures (56 mm) for both beams
have a common mechanical structure and cryostat. The cryogenic services line (QRL), which 
distributes and supplies the liquid helium to the magnets, will be installed
next to the beam line. 
 
In total 2808 bunches with a nominal intensity of $1.1\times10^{11}$ protons/bunch and a bunch-spacing of 
24.95 ns will circulate in the ring, leading to a nominal luminosity of 
$10^{34} \mathrm{cm}^{-2}\mathrm{s}^{-1}$. The total energy stored in the beam will reach
a macroscopic value of 350 MJ, which imposes severe constraints and requirements on its
safe operation, since an uncontrolled beam loss unavoidably would damage the equipment.
It is worth noting that in terms of stored energy per beam, the LHC exceeds all
previous and existing machines by a factor of 200, thus we enter unexplored territory, indeed.

\begin{figure}[htb]
\begin{center}
\begin{tabular}{cc}
\includegraphics[width=7.2cm]{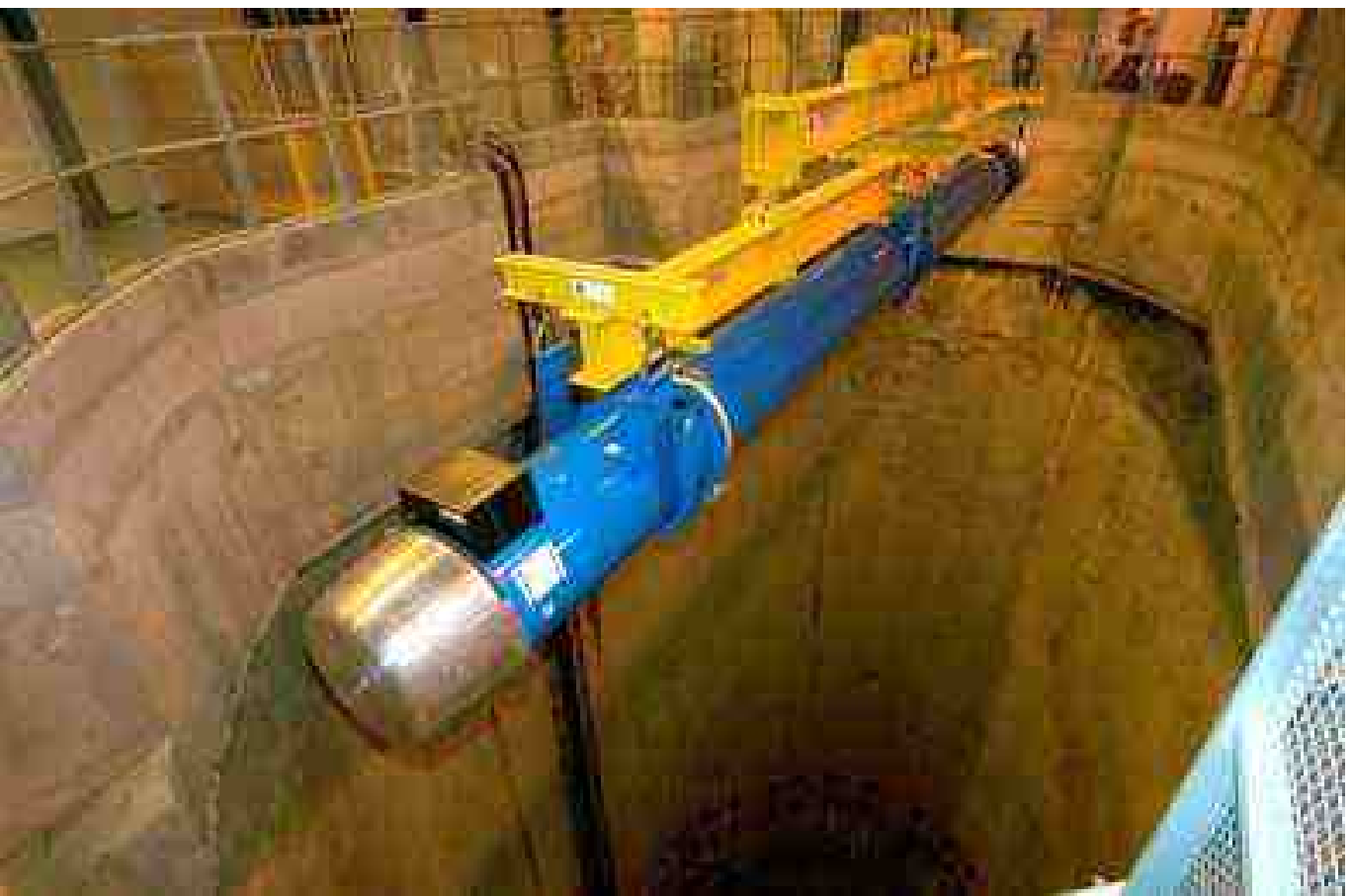} &
\includegraphics[width=7.2cm]{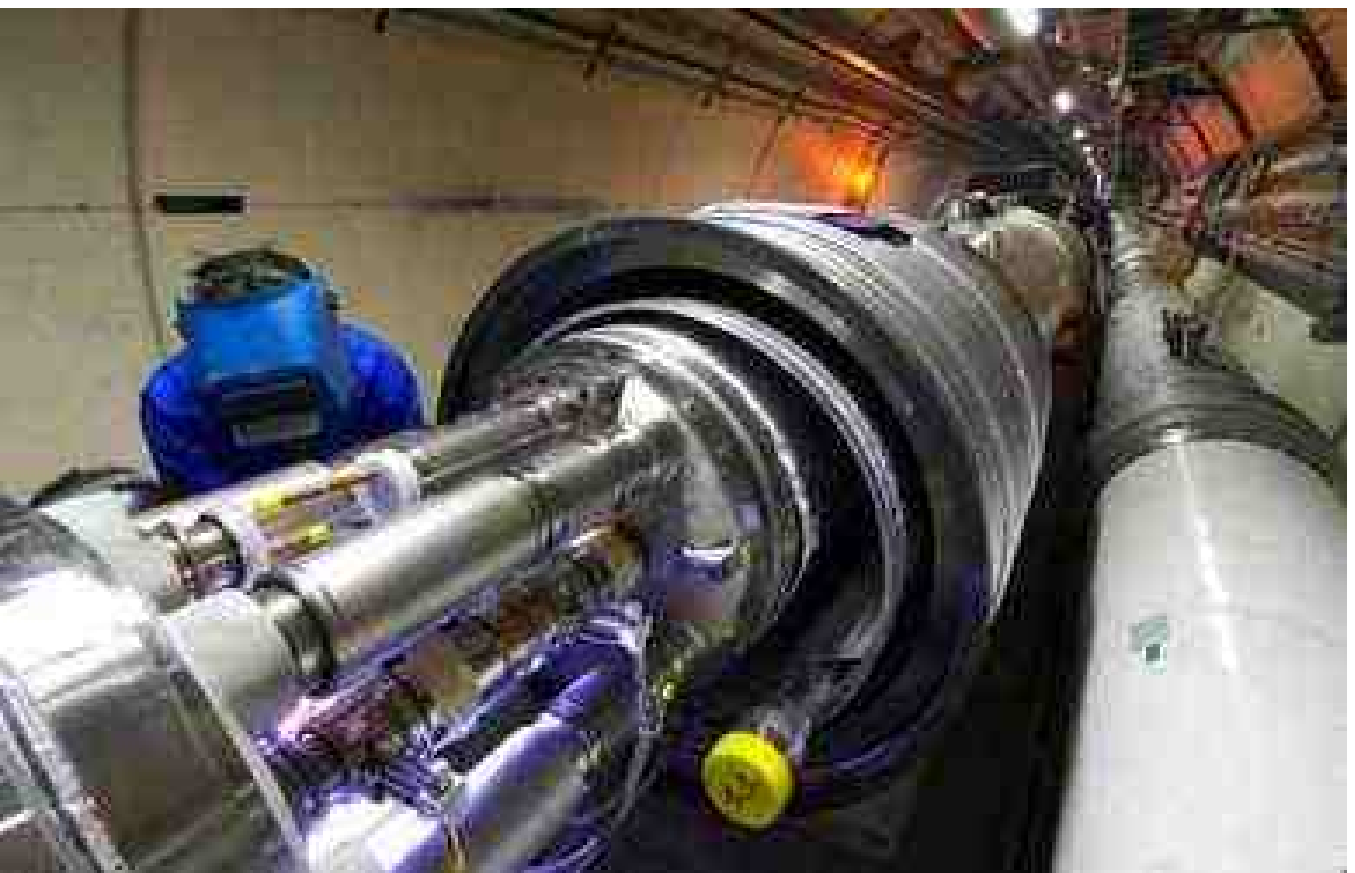}
\end{tabular}
\caption{Left~: Lowering of the first superconducting LHC dipole into the tunnel, March 2005. Right~: Interconnection
of two dipoles, next to the cryogenic services line.}
\label{LHCinstall}
\end{center}
\end{figure}

The layout of the machine consists of eight independent sectors, in order to handle in a distributed
manner the total energy of 10 GJ stored in the magnets. Being the first machine with independent sectors, 
 it presents an enormous challenge for its control and powering. Besides the four interaction points,
 where the beams will cross at a nominal angle of 285 $\mu$rad, there are warm insertion regions for
 beam dump, cleaning and acceleration.

The magnet production proceeds very well and is on schedule. To date more than 800 magnets of
excellent quality have been delivered.  The first superconducting dipole was lowered into the accelerator tunnel on Monday, $7^{\mathrm{th}}$ March (Fig.\ \ref{LHCinstall}), 
and by now about 120 dipoles are already installed in the tunnel.  
Prior to the magnets the cryogenic services line has to be
installed. This has caused problems in the past, to which CERN has reacted promptly and
successfully implemented a recovery plan. To date QRL components for four of the eight LHC sectors
have been delivered. The installation proceeds in three sectors in parallel
 and almost two sectors have been fully equipped. Recently the first QRL sub-sector has been successfully
 cooled down, after solving some minor problems which appeared during the pressure tests.
 
 The installation of the LHC in the tunnel is on the critical path for the first collisions. The LHC schedule
 \cite{LHCschedule} foresees a parallel installation of pairs of sectors, the last of them to be
 completed in June 2007. The first sector pair (sectors 7 and 8) shall be completed by May 2006 and cooled
 down for a first test with beam. This will involve beam injection from the SPS down the TI8 transfer
 line, right off point 8. The beam will pass through interaction point 8 (\LHCb) and then through 
 sector 8-7 before reaching a temporary beam dump. This will be an important system test and
 allow to pre-commission essential data acquisition and correction procedures.

%%%%%%%%%%%%%%%%%%%%%%%%%%%%%%%%%%%%%%%%

\subsection{Status of the experiments}
 \label{status-exp}

\begin{figure}[htbp]
\begin{center}
\begin{tabular}{cc}
\includegraphics[width=7.2cm]{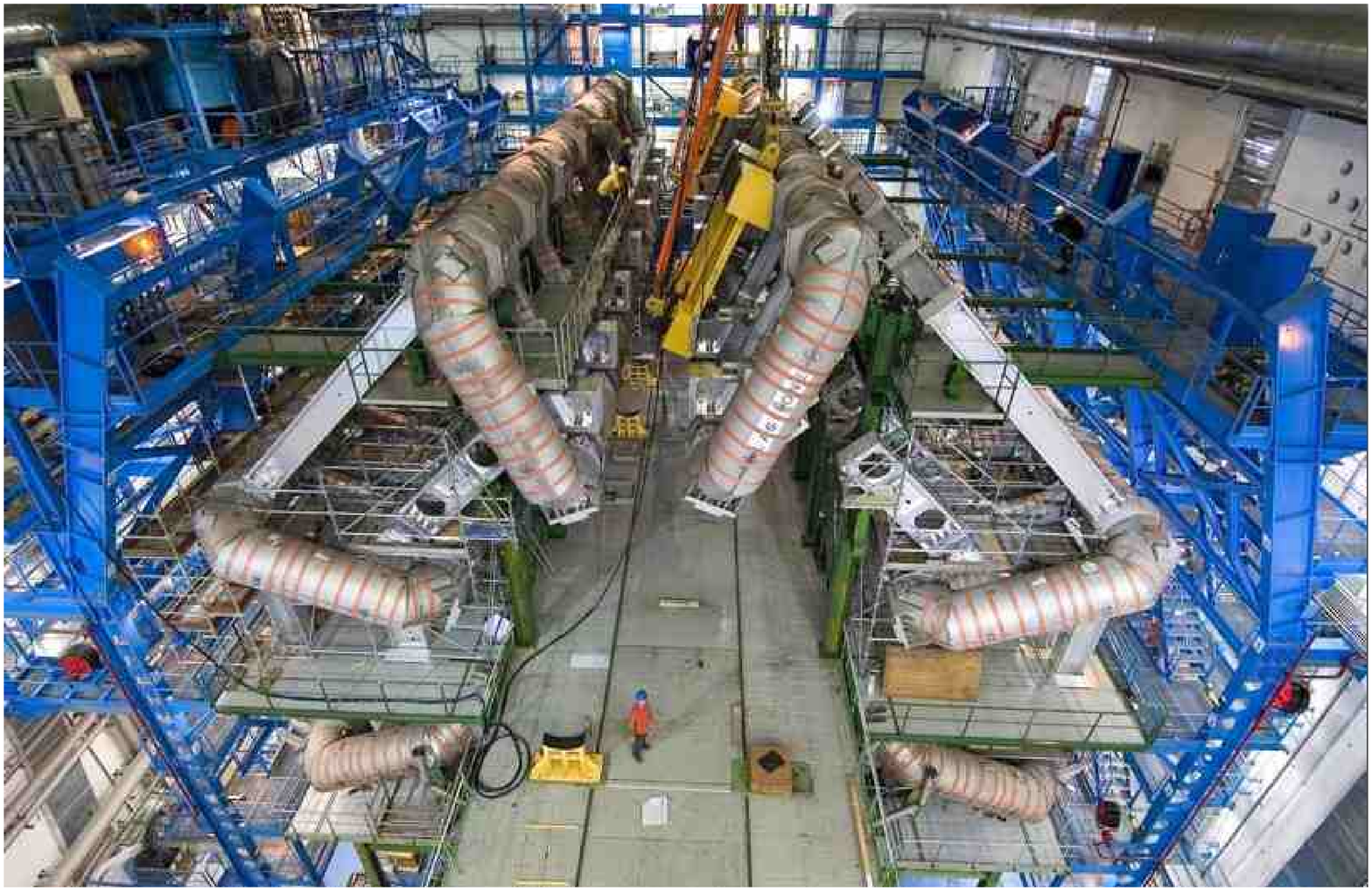} &
\includegraphics[width=7.2cm]{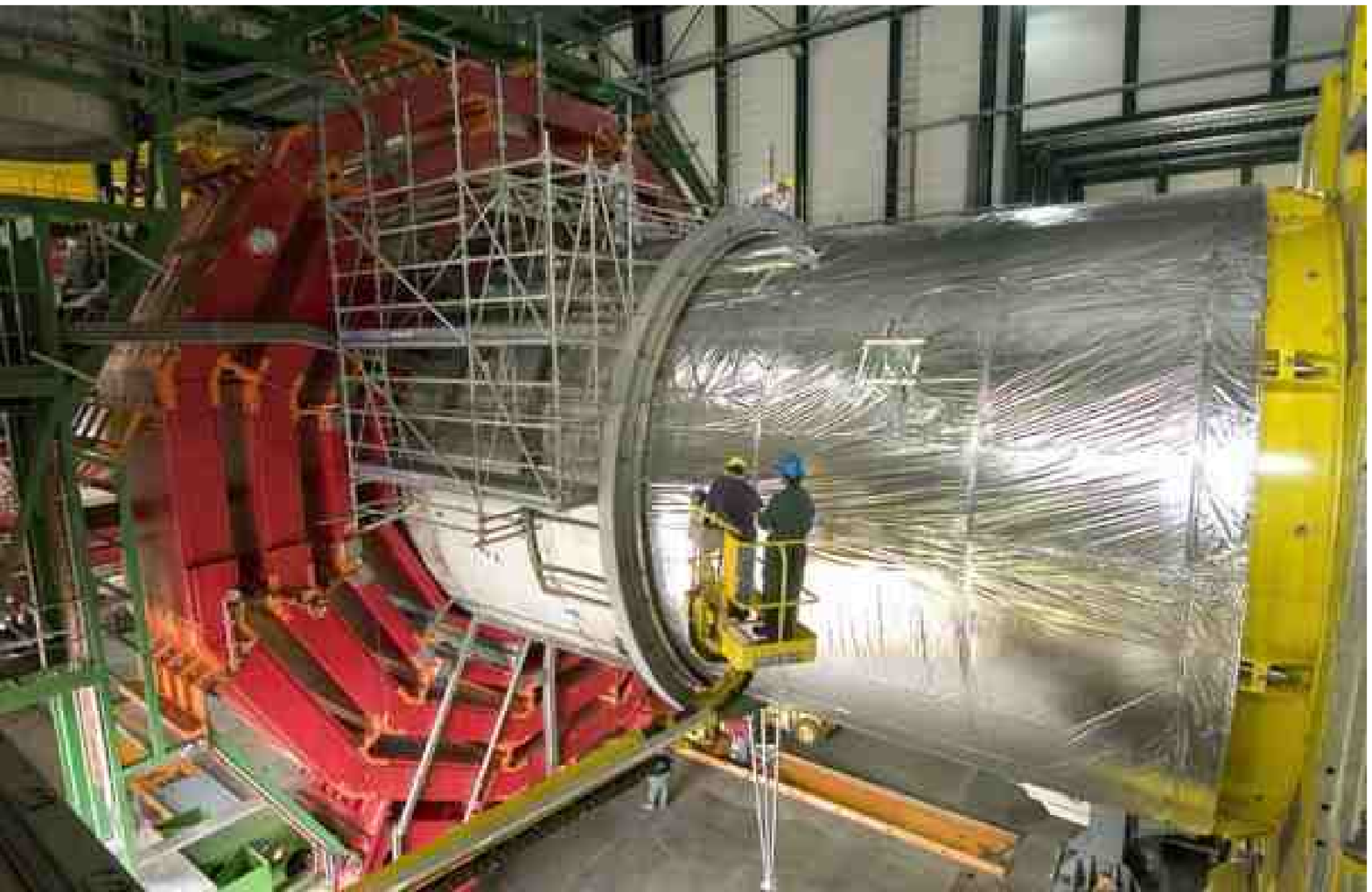}
\end{tabular}
\caption{Left~: All of the eight \ATLAS\ barrel toroid coils installed in the cavern.
               Right~: Insertion of the \CMS\ solenoid into the outer vacuum tank.}
\label{ATLASandCMS}
\end{center}
\end{figure}

The largest of the four detectors, \ATLAS\ \cite{ATLAS}, is currently being 
installed in the experimental cavern. Its enormous size
(25 m diameter, 46 m total length and 7000 tons overall weight) 
is determined by the muon system, based on air-core toroids equipped with muon chambers. At the
end of summer 2005 all eight of the barrel toroid coils have been installed (Fig.\ \ref{ATLASandCMS}, left), and the
construction and installation of the muon chambers proceeds on schedule. The construction of the electromagnetic
calorimeter, a lead/liquid argon sampling calorimeter with accordion design, is completed. It shares
the cryostat with the superconducting solenoid surrounding the central tracking system. After a successful cold
test the cryostat has been lowered into the cavern in October 2004. Also the hadronic tile calorimeter is
already underground and has registered first cosmic rays in June 2005. The planning for the construction
of the tracker, consisting of a transition radiation detector, silicon strip and pixel detectors, is tight. The
emphasis is now switching to the integration, final installation and commissioning of the sub-systems. An
important step in this direction was a combined test beam campaign in fall 2004, where parts of all sub-detectors
have been integrated and read out together, based on a common data acquisition and detector control system. 
The six-months running period has led to a good global operation experience. The common \ATLAS\ software
is  now employed to analyse a data set of about 4.5 TByte.  

The second of the two general purpose detectors, \CMS\ \cite{CMS}, is indeed compact when compared to
\ATLAS. Its overall diameter is 15 m, its length 21.6 m and its total weight 12500 tons. 
The main design difference is that it has only
one magnet system. The central superconducting solenoid, with an inner diameter of 6.32 m and a magnetic
field of 4 T, is large enough to house the tracking system, the electromagnetic and the hadronic calorimeters. 
The magnetic return yoke, made of iron rings and disks, is instrumented with muon chambers. The overall
assembly of the big mechanical parts takes place in a surface hall above the experimental cavern at point 5 of
LHC. A major milestone for the magnet
 has been achieved in summer 2005 with the final assembly of the five coil modules, its
swivelling and insertion into the outer vacuum tank (Fig.\ \ref{ATLASandCMS}, right), 
held by the central barrel wheel. The inner vacuum tank
has been inserted in November 2005. The next important steps will be the first cool-down, followed by
the magnet test in spring 2006. This will be combined with a $20^\circ$ slice test, 
where parts of all sub-detectors will be read out on cosmic ray triggers. This will be the first
trial of the \CMS\ operation procedures. The lowering of the heavy elements into the cavern will start immediately
after. Concerning the sub-detectors, the muon chamber installation is progressing very well and the
hadron calorimeter, a brass/scintillating tiles sampling calorimeter, is completed. The production of the
76848 lead-tungstate crystals for the electromagnetic calorimeter (\ECAL) is on the critical path. Its end-caps will
only be installed during the first long shutdown at the beginning of 2008. The \ECAL\ electronics integration   
is a well established procedure and test beam measurements in 2004 have shown excellent performance in terms
of energy resolution and noise. Finally, the \CMS\ silicon tracker will be the largest ever built, comprising
about $250\,\mathrm{m}^2$ of silicon sensors. 
The sensor module production will be completed in spring 2006. The complete tracker
has to be integrated at \CERN\ by November 2006 in order to be in line with the overall installation schedule.
Similar to the \ECAL\ endcaps, the pixel detector, although ready, will only be installed after the pilot run
in winter 2007-2008. 

Next to \CMS\ the \TOTEM\ experiment \cite{TOTEM} 
will be installed. It consists of CSC and GEM telescopes in the very forward direction around the
beam pipe, covering pseudo-rapidity ranges from 3.1 to 6.5, as well as of roman pots up to a distance of
220 m from the \CMS\ interaction point. The main goals of this experiment are the measurements of the
total, elastic and diffractive proton-proton cross sections, of the particle and energy flow in the very
forward direction and an absolute luminosity determination. However, in order to carry out
this physics programme, very special \LHC\ running conditions (optics) are required.

Since the b-quark production cross section peaks at the very forward direction, the \LHCb\ \cite{LHCb}
experiment is designed as a single-arm spectrometer, 
covering a pseudo-rapidity range of
$1.8 < \eta < 4.9$. The magnet, the electromagnetic and hadronic calorimeters, as well as the iron for the
muon filters have already been installed in the former \DELPHI\ cavern. Good achievement has been made
for the construction of many sub-systems, such as the vertex locator, the tracking chambers and the 
RICH detectors. \LHCb\ will collect data at a lower luminosity of 
$2\times 10^{32} \mathrm{cm}^{-2}\mathrm{s}^{-1}$. One of its essential elements, the trigger system,
has been re-organized recently, with a final storage rate of
2000 Hz after the higher level trigger stage, leading to $10^9 - 10^{10}$ B-hadrons per year. 

Finally, the \ALICE\ \cite{ALICE} detector is designed to fully exploit the heavy ion operation 
of \LHC\ (lead ions with 2.76 TeV/nucleon and an initial luminosity of  
 $10^{27} \mathrm{cm}^{-2}\mathrm{s}^{-1}$). It is installed in the former \Lthree\ cavern and re-uses
 the \Lthree\ magnet. Its main element is a huge time projection chamber of $88\, \mathrm{m}^3$,
 which will allow the reconstruction of several thousand tracks per unit of rapidity. Further
 emphasis is given to particle identification. Dedicated sub-systems cover
 restricted angular regions, such as a forward muon spectrometer based on an additional dipole
 magnet. An almost complete detector will be ready for the first proton-proton collisions in 2007.
 The full detector system will be operational for the first heavy ion run foreseen in 2008.

%%%%%%%%%%%%%%%%%%%%%%%%%%%%%%%%%%%%%%%%

\section{Commissioning and start-up scenarios}
 \label{commission}

%%%%%%%%%%%%%%%%%%%%%%%%%%%%%%%%%%%%%%%%

\subsection{LHC commissioning and early operations}
 \label{start-LHC}

In order to understand the various steps in the start-up of the LHC operations,
it is instructive to shortly review the main machine parameters and the corresponding
limitations. 
The beam energy of 7 TeV is limited by the maximal magnetic dipole field
            available and its field quality. This, on the other hand, is determined by the industrially available
            magnet technology. The chosen LHC magnet parameters (magnetic field, maximal 
            sustainable current density, temperature) leave small margins for thermal and mechanical
            stresses, which ultimately lead to quenches.

 The nominal bunch intensity (protons/bunch) is $N= 1.15\times10^{11}$, with an upper limit
            of $N = 1.7\times10^{11}$. With 2808 bunches, 
            this corresponds to a maximal beam current  of $I = 0.85\,A$.
            The limits on these parameters are imposed by several effects.
             Beam-beam effects, which lead to a tune ($Q$) spread, have to be minimized in order to avoid
             resonances, since the allowed region in $Q$-space is rather restricted. Resonances are also
             avoided by an excellent magnetic field quality, by correction circuits and optimal feedback from the
             beam instrumentation. A high operation efficiency and thus the maximal achievable integrated
             luminosity is obtained by minimizing quenches and beam aborts. Here collimators and the cleaning 
             insertions play an important role. However, the sustainable radiation dose in the cleaning insertions
             is just compatible with the nominal beam intensities. Also the enormous
               energy stored in the beams represents a serious damage potential and therefore has to be well
              understood and controlled. The high beam intensities, coupled to badly conducting
              collimator materials (eg.\ graphite) induce large wake fields which cause collective beam instabilities.
              This effect can be controlled by first limiting the beam intensity, and second by a proper choice of
              collimators. In a first phase, graphite collimators will be used, thanks to their robustness and thus
              increased machine safety. Since with this choice the beam current is limited to $I<0.3\,A$, 
              in a second phase they
              will be replaced by copper collimators, allowing for $I<0.85\,A$. These
              are good conductors, but would be seriously damaged in case of a full beam impact.
              Finally, the electron cloud effect \cite{ElectronCloud} puts heat load on the beam screen, which increases
              for smaller bunch spacing. Recent studies show that the nominal number of bunches allows for
              a stable operation, if there is a prior conditioning of the surfaces by a so-called beam scrubbing.
              Initially this effect can be more easily controlled by an increased bunch spacing, for example 75 ns.
 
 The physical beam size, $\sigma = \sqrt{\beta \epsilon}$, is determined by the machine's
              $\beta$-function and the emittance $\epsilon = \epsilon_\mathrm{n}/\gamma$, where
              $\gamma$ is the Lorentz factor and 
              $\epsilon_\mathrm{n}$ the so-called normalized emittance. The latter is limited to 
              $\epsilon_\mathrm{n} < 3.75\, \mu$m by the injector chain and the main dipole aperture. The luminosity
              at the interaction points (IPs) increases with smaller beam size $\sigma^*$, 
              thus a smaller $\beta$-function. Its value at an IP is denominated $\beta^*$. A limit on this
              parameter ($\beta^* > 0.55$ m) is imposed by the physical aperture of the first triplet of
              superconducting quadrupole magnets around the IPs. The $\beta$-function in proximity of an IP
              can be approximated by $\beta(s) \approx \beta^* + s^2/\beta^*$, $s$ being the distance from the IP. 
              Taking a nominal $\sigma^* = 16.6\,\mu$m and a distance to the first quadrupoles of $\approx 23\,$m,
              the beam size at the quadrupoles is of the order of a millimetre, which starts to be of a similar
              order of magnitude as the quadrupole aperture. 

For nominal bunch intensities and spacing the beam-beam effects near an IP can be reduced by
               introducing a finite beam crossing angle, $\approx 300\, \mu$rad. This crossing angle is again
               limited by the aperture of the nearest quadrupole magnets and the corresponding stress on these
               magnets and the collimators.                

All the above mentioned issues have to be taken into account in order to achieve the main objective
of the early LHC operation, which is to establish colliding beams as quickly as possible, safely and without
compromising further progress. The general approach will be to initially take two multi-bunch beams with
moderate intensity to high energy and collide them at zero crossing angle. Graphite collimators will be
installed at the beginning.

Currently the hardware commissioning, system tests, machine and transfer lines checkout are foreseen
for late spring to early summer 2007. Then the commissioning with beam will start, followed by a pilot
run in fall 2007. The start-up of the machine is planned in four stages, approaching gradually the ultimate
machine parameters. In a first stage the LHC will run with $43\,\times\,43$ bunches, moving to
 $156\,\times\,156$ bunches with moderate intensities ($N\approx 3\times10^{10}$), zero to partial
 squeeze and zero crossing angle. In a second stage with $936\,\times\,936$ bunches and partial 
 squeeze a luminosity of up to $4\times10^{32}\mathrm{cm}^{-2}\mathrm{s}^{-1}$ should be reached.
 The third stage would correspond to the start of the 25 ns operations, with intensities up to $N=5\times10^{10}$
 and almost full squeeze. During this stage a luminosity of $2\times10^{33}\mathrm{cm}^{-2}\mathrm{s}^{-1}$
 will be approached, which currently is also assumed in the physics studies for the first year(s) of LHC 
 operation. The fourth stage, when all parameters will be pushed to the nominal values 
 (luminosity of $10^{34}\mathrm{cm}^{-2}\mathrm{s}^{-1}$), will only be reached after a few years.
 It necessitates the exchange of the graphite
 with copper collimators and the installation of the complete beam dump system. Since it is not clear how
 fast the first operation stages will be passed through and what the machine operation and detector efficiencies
 will be, it is very difficult to give a precise estimate of the integrated luminosity on tape at the end of
 2008, after the first long physics run. Based on the numbers above, this could be anywhere in the
 range of 0.1 to 10 $\mathrm{fb}^{-1}$ \cite{GianottiMangano}.

%%%%%%%%%%%%%%%%%%%%%%%%%%%%%%%%%%%%%%%%

\subsection{Commissioning of the experiments}
 \label{start-exp}

All experiments are expected to be ready by end of June 2007. However, parts or all of the
installed sub-detectors can be commissioned and pre-calibrated already well before we
have first collisions in LHC. Cosmic ray muons will be used by all detectors in order to obtain
initial alignment and calibration constants for the barrel parts mainly. These muons are also
very useful for debugging and mapping of dead-channels. An estimate of the rate
is approximately 1 - 5 kHz for muons with an energy at the surface exceeding 10 GeV. Out of
these a few Hz might be useful for calibration. It is worth repeating that some sub-detector systems 
have already been commissioned with cosmic muons in 2005, and \CMS\ foresees a full
system test at the surface, with the magnetic field on, in 2006.

Once at least one beam is circulating, beam-halo muons will traverse the experiments and thus
allow for further alignment and calibration efforts, now with emphasis on the end-caps of the detectors. The rate
for muons with an energy above 100 GeV is estimated to about 1 kHz. At the same time, beam-gas
events will be registered, which already resemble the later proton-proton collisions. However,
the spectrum of the produced tracks is much softer, with typically $p_\mathrm{T} < 2$ GeV/$c$. 
Nevertheless, with a rate of 25 Hz for reconstructed tracks with  $p_\mathrm{T} > 1$ GeV/$c$, coming
from a vertex with $|z| < 20\,$cm, it might be possible to obtain a first alignment of the inner
trackers to about $100\,\mu$m.

Finally, with the first collisions in hand, the trigger and data acquisition systems will be timed-in,
the data coherence checked, sub-systems synchronized and reconstruction algorithms debugged
and calibrated. The electromagnetic and hadronic calorimeters will be calibrated with first
physics events. For example, the initial crystal inter-calibration precision of about 4\%
for the \CMS\ \ECAL\ will be improved to about 2\% by using the $\phi$-symmetry of the energy
deposition in minimum-bias and jet events. Later the ultimate precision ($\approx 0.5\%$) and the absolute
calibration will be obtained using $\mathrm{Z}\rightarrow \mathrm{e}^+ \mathrm{e}^-$ decays and
the $E/p$ measurements for isolated electrons, such as in $\mathrm{W}\rightarrow \mathrm{e}\nu$ decays
\cite{ECALCalibration}.
The latter requires a well understood tracking system. The uniformity of the 
hadronic calorimeters can be checked with single pions and QCD jets. In order to obtain the
jet energy scale to a few per-cent or better, physics processes such as 
$\mathrm{Z}(\rightarrow \ell\ell) + \mathrm{jet}$ or $\mathrm{W}\rightarrow$ 2 jets in top pair events
will be analyzed.
Finally, the tracker and muon system alignment will be carried out with generic tracks, isolated muons or
$\mathrm{Z}\rightarrow \mu^+ \mu^-$ decays. Regarding all these calibration
and alignment efforts, the ultimate statistical precision should be 
achieved after a few days of operation in most cases. Then systematic effects
have to be faced, which, eg., implies that pushing the tracker $R\phi$ alignment from an initial
$100\,\mu$m to about $5\,\mu$m might involve at least one year of data taking. A more detailed
review of the initial detectors and their performance can be found in Ref.\ \cite{GianottiMangano}.

%%%%%%%%%%%%%%%%%%%%%%%%%%%%%%%%%%%%%%%%

\section{Early Physics}
 \label{physics-early}

The very early goals to be pursued by the experiments, once the first data are on tape, are
three-fold~: (a) It will be of utmost importance to commission and calibrate the detectors
in situ, with physics processes as outline above. The trigger performance has to be understood
in a possibly unbiased manner, by analyzing the trigger rates of minimum-bias events,
QCD jet events for various thresholds, single and di-lepton as well as single and di-photon
events. (b) It will be necessary to measure the main SM processes (cf.\ section \ref{SM-tests}) and
(c) prepare the road for possible discoveries (section \ref{discoveries}). 

It is instructive to recall the event statistics collected for different types of processes. 
For an integrated luminosity of $10\,\mathrm{fb}^{-1}$ per experiment, we expect about $10^8$
$\mathrm{W}\rightarrow \mathrm{e}\nu$ events on tape, a factor of ten less $\mathrm{Z}\rightarrow \mathrm{e}^+ \mathrm{e}^-$ 
and some $10^6$ $\mathrm{t}\bar\mathrm{t}\rightarrow \mu + X$ events. Even if a trigger bandwidth of only 10\%
is assumed for  QCD jets with $p_\mathrm{T} > 150$ GeV/$c$,  $\mathrm{b}\bar\mathrm{b}\rightarrow \mu + X$ and minimum-bias events, 
it still gives about $10^7$ events on tape for each of these channels. Also the existence
of supersymmetric particles,
for example gluinos with $m_{\tilde\mathrm{g}}\approx 1$ TeV/$c^2$,  or a
Higgs with $m_\mathrm{H} \approx 130$ GeV/$c^2$,
would result in sizeable events statistics ($10^3 - 10^4$).
Summing up everything, we estimate some $10^7$ events to tape
every three days. Integrated over a full year, this
amounts to 1 PByte of data per experiment. Apart the computing challenge to be faced for the data
storage, distribution, reconstruction and analysis, this means that the statistical uncertainties
will be negligible after a few days, for most of the physics cases. The analyses results will be dominated
by systematic uncertainties, be it the detailed understanding of the detector response, theoretical
uncertainties or the uncertainty from the luminosity measurements. A very detailed review
of the early physics cases and the related analysis issues can be found in Ref.\ \cite{GianottiMangano}.

%%%%%%%%%%%%%%%%%%%%%%%%%%%%%%%%%%%%%%%%

\subsection{Tests of the Standard Model}
 \label{SM-tests}

There are many good reasons to investigate considerable efforts in the measurements
of SM processes. We are sure that these have to be seen and thus they can serve
as a proof for a working detector (a necessary requirement before any claim of discovery is made).
Above we have mentioned that some SM processes are excellent tools to calibrate
parts of the detector. However, such measurements are also interesting in their own right. 
We will be able to challenge the SM predictions at unprecedented energy and momentum transfer scales,
by measuring cross sections and event features for minimum-bias events, QCD jet production,
W and Z production with their leptonic decays, as well as top quark production.
This will allow to check the validity of the Monte Carlo generators, both at the highest energy scales 
and at small momentum transfers, such as in models for the omnipresent underlying event. 
The parton distribution functions (pdfs) can be further constrained or measured for the first time in
kinematic ranges not accessible at \HERA. Important tools for pdf studies will be 
jet+photon production or Drell-Yan processes. Of course, SM processes are
backgrounds for the new physics searches. In particular W/Z+jets, QCD multi-jet and top pair 
production will be important backgrounds to a large number of searches and therefore have to
be understood in detail. In the following two examples are discussed a bit further.

\begin{figure}[t]
\begin{center}
\begin{tabular}{cc}
\includegraphics[width=7.7cm]{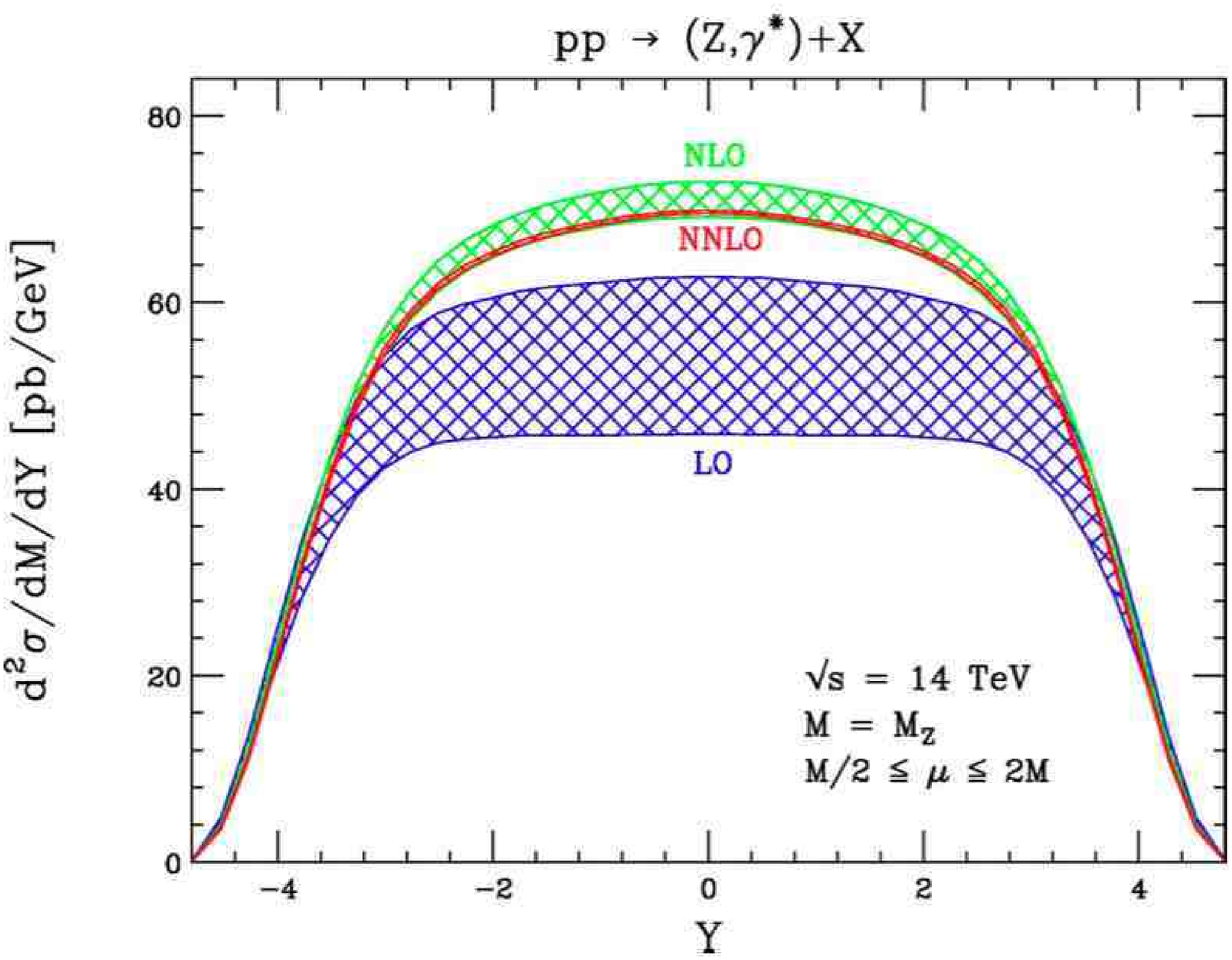} &
\includegraphics[width=6.8cm]{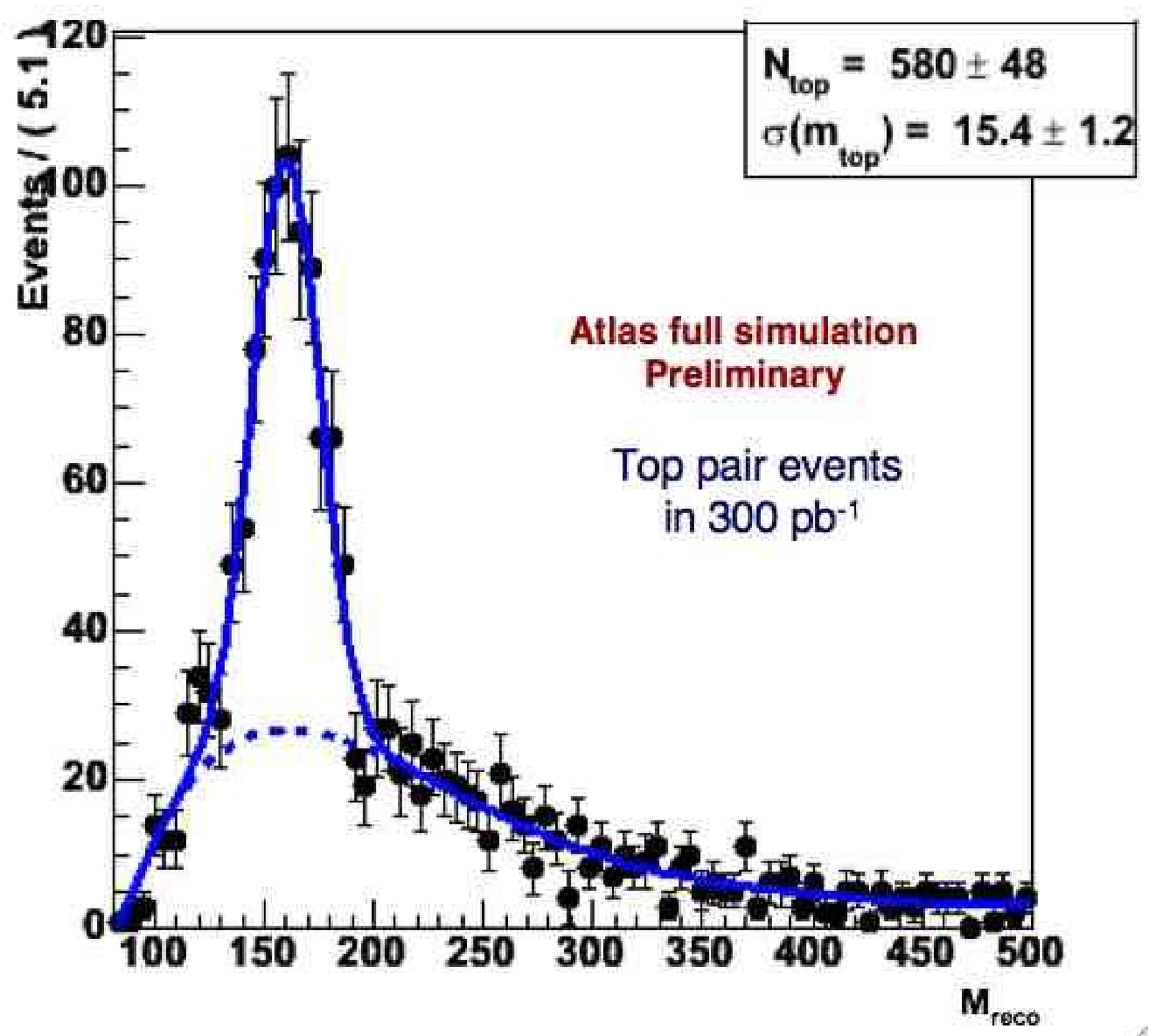} 
\end{tabular}
\caption{Left~: Predictions for the rapidity dependence of Z boson production at \LHC, 
at various orders in QCD perturbation theory \cite{Anastasiou2004}. Right~: \ATLAS\ simulation of the 
top quark mass reconstruction for an integrated luminosity of $300\,\mathrm{pb}^{-1}$ \cite{ATLAStop}. }
\label{SMplots}
\end{center}
\end{figure}

Most likely the theoretically best known cross section at \LHC\ will be for
 lepton pair production, via the Drell-Yan production of W and Z bosons. In 2004 the first differential 
next-to-next-to-leading order (NNLO) QCD calculation for vector boson production 
in hadron collisions was completed by Anastasiou et al.\ \cite{Anastasiou2004}. This group has
calculated the rapidity dependence for W and Z production at NNLO (Fig.\ \ref{SMplots}, left). They have
shown that the perturbative expansion stabilizes at this order in perturbation theory
and that the renormalization and factorization scale uncertainties are  drastically 
reduced, down to the level of one per-cent. Recent studies \cite{HERALHCws} conclude
that the dominant uncertainties will be related to the knowledge of the pdfs,
currently  estimated at the 4-5\% level. This pdf uncertainty, as well as experimental and luminosity uncertainties,
can be considerably reduced or completely eliminated by looking at ratios of cross sections, such
as the rapidity dependence of $\mathrm{W}^+/\mathrm{W}^-$. On the other hand, the rapidity dependence of 
vector boson production will impose important constraints on the available pdf sets. Finally, 
in Ref.\ \cite{DittmarLumi} it has been shown that the Drell-Yan process will be an important alternative tool
for the determination of the machine and parton luminosities. Again, if normalized to the Drell-Yan production
of, eg., Z bosons, many other SM processes can be predicted with considerably reduced uncertainties and
without the knowledge of the machine luminosity.

\LHC\ will be a top factory and thus offer a very rich top physics programme. Basically top quark production
will be seen immediately. As a recent \ATLAS\ study shows \cite{ATLAStop}, even with a simple
selection a very clear peak in the distribution of the reconstructed top mass is obtained with very
small integrated luminosity (Fig.\ \ref{SMplots}, right). The selection only requires missing transverse
energy, one high-$p_\mathrm{T}$ and isolated lepton, at least four jets and a cut on the reconstructed
hadronic W mass, but no b-tag. From this it is clear that, contrary to the \TEVATRON, top production
will be an important calibration tool, both for the jet energy scale by looking at the hadronically reconstructed
W mass and for the b-tagging algorithms. Some of the physics topics to be addressed will be the measurements
of top quark properties, such as its production and decay probabilities, its couplings, spin and mass.
For the latter an ultimate precision of 1 GeV/$c^2$ is claimed to be achieved.

%%%%%%%%%%%%%%%%%%%%%%%%%%%%%%%%%%%%%%%%

\subsection{Early discovery scenarios}
 \label{discoveries}

Regarding possible scenarios of discoveries to be made with the first year's \LHC\ data,
we summarize a study published in Ref.\ \cite{GianottiMangano}. There three cases have been
identified according to the experimental and theoretical difficulties to be faced in order to 
have an unambiguous claim of discovery~: 
(a) An "easy" case would be the appearance of a new heavy resonance which
decays into electron or muon pairs; (b) an intermediate case could be the search for supersymmetric
particles; (c) the discovery of a very light Higgs ($m_\mathrm{H} \approx 120\,$ GeV/$c^2$) is considered
to be a difficult case.  Whatsoever scenario is realized in nature, it is clear that the \LHC\ is 
a perfect place to look for evidence of new physics in the TeV energy range, thanks to the large
phase space and the large rates for new particles production predicted by many of the SM extensions.

\begin{figure}[htbp]
\begin{center}
\begin{tabular}{cc}
\includegraphics[width=7.2cm]{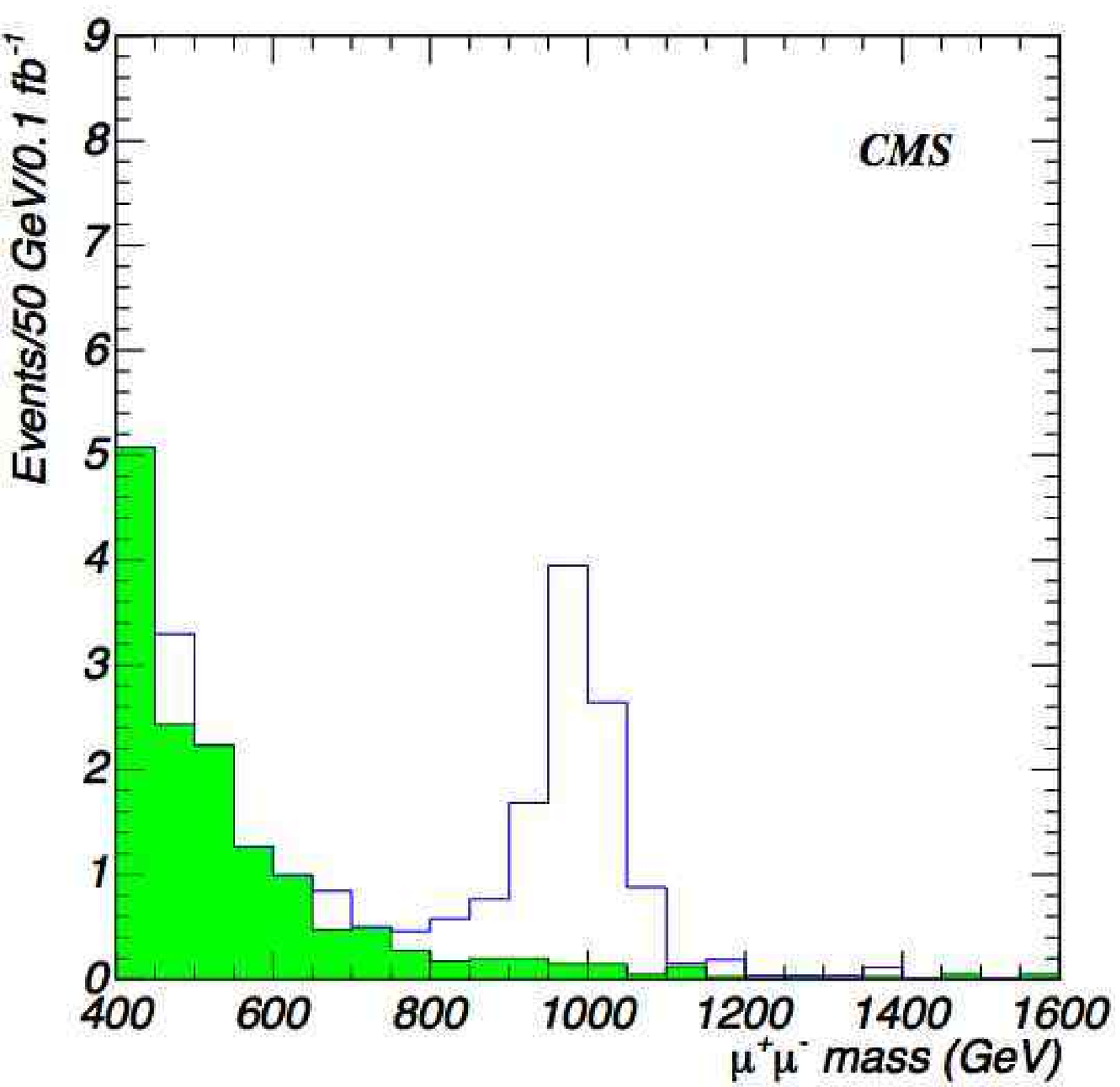} &
\includegraphics[width=7.2cm]{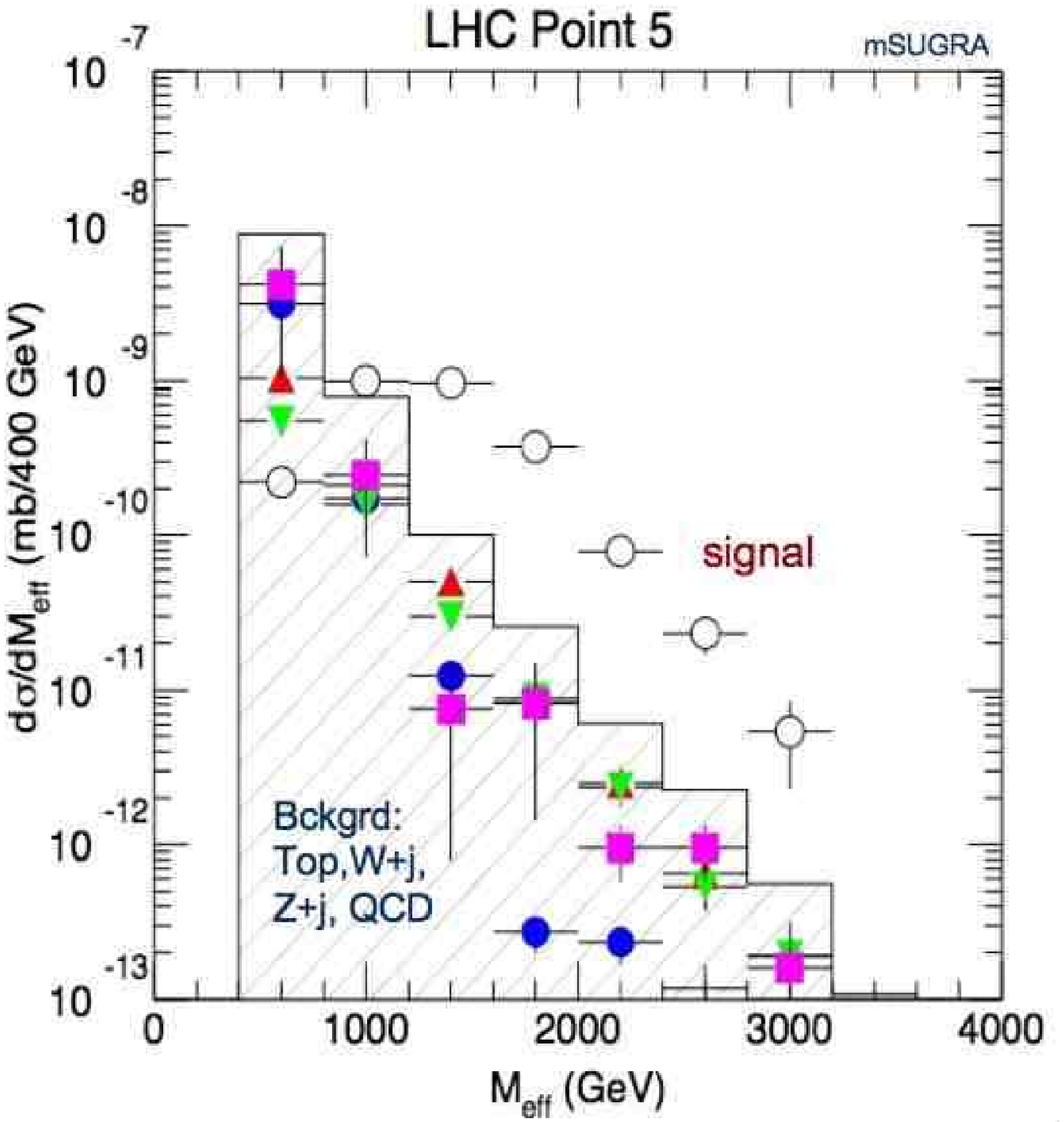} 
\end{tabular}
\caption{Left~: Simulated appearance of a $\mathrm{Z}_\psi$ resonance in the invariant mass spectrum of muon pairs measured
in \CMS\ \cite{CMSCousins}. Right~: Reconstructed effective mass (see text for its definition) for various SM backgrounds and a SUSY signal as predicted for a particular point in the mSUGRA parameter space. 
The results have been obtained with a
simulation of the \ATLAS\ detector \cite{ATLASPTDR}.}
\label{SusyPlots}
\end{center}
\end{figure}

First we discuss the "easy" case. 
We use the notation Z' for any generic new heavy gauge boson with a mass up to several TeV/$c^2$.
Such heavy gauge bosons appear in models with extensions of the SM gauge group, such as the
recent Little Higgs models \cite{LittleHiggs}, in theories with dynamical electro-weak symmetry breaking or
generally in Grand Unified Theories (GUTs) (see Ref.\ \cite{ZprimeSummary} for an overview of Z' searches). 
If such a heavy gauge boson has SM-like couplings to
leptons and quarks, we could expect sizeable production cross sections and branching ratios to
$\mathrm{e}^+\mathrm{e}^-$ and   $\mu^+\mu^-$, resulting in a very clear signature  
above a low and well understood background. The decay leptons would have very high transverse momentum
and be isolated, thus easy to be triggered on. 
Figure \ref{SusyPlots} (left) shows the result of a \CMS\ study \cite{CMSCousins} for
the detection of a $\mathrm{Z}_\psi$ \cite{Zpsi} with a mass of 1 TeV/$c^2$ via its decay into muon pairs, after trigger and
offline reconstruction. Already with
less than $1\,\mathrm{fb}^{-1}$, such a signal cannot be missed. A similar study by \ATLAS\ concludes
that a Z', predicted by a Sequential SM \cite{SSM}, could be detected with about $1.5\,\mathrm{fb}^{-1}$
up to masses of 2 TeV/$c^2$, via its decay to $\mathrm{e}^+\mathrm{e}^-$. The necessary energy calibration and  
understanding of the lepton identification efficiency will be obtained from the processes 
$\mathrm{Z}\rightarrow\ell\ell+\,$jet and Drell-Yan Z production, measured on the real data.

Heavy resonances also appear in models with extra dimensions. A particular realization of the Randall-Sundrum (RS)
model with massive Kaluza-Klein excitations (gravitons) around 1 TeV has been analyzed using the full detector simulation and 
reconstruction of \CMS\ \cite{CMSRS}. For sizeable couplings to electrons such a graviton would result
in a very clear peak in the invariant mass distribution of high-$p_\mathrm{T}$ and isolated electron-positron
pairs, over a very small background, allowing an unambiguous discovery with  $10\,\mathrm{fb}^{-1}$ of
collected data. An important experimental aspect, which has been considered here, is that electrons in the
TeV energy range may lead to saturation effects in the readout of the electromagnetic calorimeters. 
Once the existence of such a heavy resonance is established, it will be most exciting to analyse the further
data and discriminate between the various models which predict such an object.
Possible approaches will be to measure forward-backward asymmetries in the case of a Z' \cite{DittmarGiolo} 
or to look for other hints of extra dimensions such as large missing energy or very energetic isolated photons.

A large fraction of the analyses for an early discovery at \LHC\ concentrate on supersymmetric extensions of the 
SM. Because of the extremely large parameter space, some specific benchmark models and points in parameter space
have been chosen \cite{Benchmarks} for a more in-depth study of the discovery potential, such as 
mSUGRA with its minimal set of five parameters (for an introduction see
\cite{SUSY} and references therein). Among these are the universal scalar ($m_0$) 
and gaugino ($m_{1/2}$)  masses, which are fixed at the GUT scale. The full spectrum of supersymmetric particles
at the TeV scale is then derived by simply employing the renormalization group equations. Over a considerable
region of the parameter space, the production cross sections for supersymmetric partners of quarks (squarks)
and gluons (gluinos) are very large, thanks to their strong (QCD) couplings to the incoming partons. As an
example, up to 100 events per day would be expected for gluino and squark masses of $\sim 1$ TeV/$c^2$, at
a luminosity of $10^{33} \mathrm{cm}^{-2}\mathrm{s}^{-1}$. The subsequent decays, which typically
occur via cascades to lighter supersymmetric (eg.\ charginos, neutralinos) and SM particles, lead to spectacular
experimental signatures with very characteristic topologies. These are triggered on by looking for high-energetic
multi-jet and multi-lepton events. If $R$-parity is conserved, the lightest SUSY particle (a neutralino in most models)
would escape undetected. Therefore one of the most important characteristics (and trigger conditions) 
of such events would be large
missing energy, $/\!\!\!\!\!E_\mathrm{T}$. A general approach of SUSY searches consists in the analysis of
topological variables, such as the effective mass $M_\mathrm{eff} = /\!\!\!\!\!E_\mathrm{T} + \sum p_\mathrm{T}^\mathrm{jet}$,
cf.\ Fig.\ \ref{SusyPlots} (right). A SUSY signal is expected to appear as an excess over the SM backgrounds
(top production, W/Z+jets, QCD multi-jet events) in the large $M_\mathrm{eff}$ region. However, in order to have
clear evidence of a signal, two important aspects have to be considered. First, a good experimental understanding
and calibration of the $/\!\!\!\!\!E_\mathrm{T}$ measurement is required. Second, a good theoretical control of the
many SM backgrounds is not trivial to achieve with the current predictions and Monte Carlo models at hand. The problem
is that multi-jet events, particularly in the high-$E_\mathrm{T}$ tail, are known to be badly simulated by the
widely used parton shower models. The incorporation of matrix element corrections is absolutely essential for
a reliable prediction \cite{GianottiMangano}. Various approaches in this direction have appeared recently
\cite{MCGenerators}, but still a lot of effort is needed in order to reach a mature
level of understanding. Therefore an early SUSY discovery via a topological search is considered to be of
"intermediate" difficulty. However, if the above-mentioned issues are under control, squarks with masses
up to 1.5 (2) TeV/$c^2$ will be discovered with only 1 (10) $\mathrm{fb}^{-1}$ of data on tape.

The electro-weak fits in the context of the SM, 
combined with the limits from the direct \LEP\ searches, indicate that we can expect a Higgs
boson with $114 < m_\mathrm{H} < 219$ GeV/$c^2$ \cite{ewfits}, with a preference for a very light Higgs,
$m_\mathrm{H} \approx 120$ GeV/$c^2$. In this mass range the dominant production mode is gluon fusion,
but additional channels such as vector boson fusion (VBF) and associated Higgs production (WH, ZH, $\mathrm{t}\bar\mathrm{t}$H)
are also considered. A very light Higgs dominantly decays  to $\mathrm{b}\bar\mathrm{b}$ pairs. Because
of the enormously large QCD backgrounds, a detection in this decay channel might only be possible
in the $\mathrm{t}\bar\mathrm{t}$H case. However, this will be extremely difficult, both because of experimental issues
(jet energy scale, combinatorics, b-tagging) and the theoretical control of the backgrounds, such as  $\mathrm{t}\bar\mathrm{t}+$jets.
The channel $\mathrm{H}\rightarrow\tau\tau$ is investigated for VBF, which requires an excellent experimental
understanding of very forward jets, a central jet veto and a $\tau$ trigger with good efficiency. Although their branching ratios are
at the per-cent or even per-mille level, the channels $\mathrm{H}\rightarrow\mathrm{Z}\mathrm{Z}^*\rightarrow4\ell$
($\ell = \mathrm{e},\mu$) and $\mathrm{H}\rightarrow\gamma\gamma$ are the most promising ones. In both
cases the experimental signature would be a clear mass peak over a rather smooth background. 
In the former case good momentum resolution and understanding of the lepton isolation efficiency is required.
The  $\mathrm{H}\rightarrow\gamma\gamma$ case has been the benchmark channel for the design
of the electromagnetic calorimeters, since a detection above the very large background is only possible
with a mass resolution at the 1\% level. This channel is studied for inclusive production, but also in the context
of VBF and associated Higgs production. For a Higgs mass between 140 and 180 GeV/$c^2$, the
most promising discovery channel turns out to be $\mathrm{H}\rightarrow\mathrm{W}\mathrm{W}^*\rightarrow2\ell\,2\nu$,
although a mass peak reconstruction is not possible because of the neutrinos. Nevertheless, with a jet veto and cuts related
to the lepton kinematics a signal over background ratio above unity is achieved and
a discovery should be in reach with the first few $\mathrm{fb}^{-1}$ of data \cite{DittmarHiggs}. 

In summary, in the context of the SM 
a Higgs with a mass around 120 GeV/$c^2$ can be detected in the first $10\,\mathrm{fb}^{-1}$ of data
only if several channels (and the experiments) are combined. Each of the channels
$\mathrm{H}\rightarrow\gamma\gamma$, $\mathrm{t}\bar\mathrm{t}\mathrm{H}\rightarrow\mathrm{b}\ell\nu\,
\mathrm{bjj}\,\mathrm{bb}$ and $\mathrm{qqH}\rightarrow\mathrm{qq}\tau\tau$ might contribute some $2\sigma$ of
total significance per experiment. Thus an observation of all channels is important. This requires 
well functioning experiments, due to the many different requirements on the detector performance, as well
as a good theoretical understanding of the backgrounds 
($\mathrm{t}\bar\mathrm{t}\mathrm{j}$, $\mathrm{t}\bar\mathrm{t}\mathrm{jj}$,
W/Z+jets, QCD multi-jets, prompt photon production) at the 10\% level or better. It is worth noting that
considerable progress has been made recently in the calculations of higher order QCD corrections 
and $p_\mathrm{T}^\mathrm{H}$ resummation for Higgs production, see eg.\ Refs.\ \cite{BabisHiggs, Grazzini}.
Also the case of supersymmetric Higgs bosons is investigated, in particular for large
$\tan\beta$ (the ratio of the vacuum expectation values of the two Higgs doublets) when the Higgs coupling
to $\mathrm{b}\bar\mathrm{b}$ and $\tau^+\tau^-$ is enhanced. 
Summaries of the discovery potentials for various SUSY parameter regions can be found in
Refs.\ \cite{CMSHiggs} and \cite{Carena}.

%%%%%%%%%%%%%%%%%%%%%%%%%%%%%%%%%%%%%%%%

\subsection{The relevance of B-Physics}
 \label{Bphysics}

Heavy flavour and in particular B-physics will be part of the toolkit to look for
new physics at \LHC. It will be possible to over-constrain the CKM matrix in many
ways, to compare tree-level dominated processes to processes which involve
penguin or box diagrams, and thus to reveal new (CP violating) physics in case 
discrepancies are found. The CKM-angle $\gamma$ will be measured
by \LHCb\ in various channels, with different sensitivity to new physics. For example,
the time-dependent CP asymmetry of 
$\mathrm{B}_\mathrm{s}\rightarrow\mathrm{D}^-_\mathrm{s}\,\mathrm{K}^+$ depends
on $\gamma$ already at the tree level, in contrast to 
$\mathrm{B}^0\rightarrow\pi^+\pi^-$ and $\mathrm{B}_\mathrm{s}\rightarrow\mathrm{K}^+\mathrm{K}^-$.
A comparison of the asymmetries in these channels, as well as of the various decay
rates in the $\mathrm{B}^0\rightarrow\mathrm{D}^0\,\mathrm{K}^{*0}$ system will give further sensitivity. A special
advantage of \LHC\ compared to B-factories is the copious production of  $\mathrm{B}_\mathrm{s}$ mesons.
The $\mathrm{b}\rightarrow\mathrm{s}$ transition is  an interesting candidate for the appearance of new 
physics. Rare decays could reveal new phenomena up to the TeV energy scale and complement
the direct searches. Deviations from the SM predictions may be seen in the 
$\mathrm{B}_\mathrm{s}\rightarrow\phi\phi$ decay and in some regions of the SUSY parameter space the branching ratios for 
$\mathrm{B}_\mathrm{s}\rightarrow\mu^+\mu^-$, $\mathrm{B}_\mathrm{d}\rightarrow\mathrm{K}^*\mu^+\mu^-$
and $\mathrm{B}_\mathrm{s}\rightarrow\phi\mu^+\mu^-$ are considerably enhanced above
the SM values. Finally, the $\mathrm{B}_\mathrm{s}$ mixing and oscillation may be affected. This could be observed
by a larger than expected CP violation in $\mathrm{B}_\mathrm{s}\rightarrow\mathrm{J}/\psi \phi$ or by an
oscillation frequency $\Delta m_\mathrm{s}$ which is higher than the SM expectation (around $20\,\mathrm{ps}^{-1}$).
The \LHCb\ experiment is very well designed to measure the $\Delta m_\mathrm{s}$ parameter, thanks to its dedicated 
triggers, good particle identification capabilities and 
excellent proper time resolution (40 fs). This allows to achieve
a $\Delta m_\mathrm{s}$ reach of $68\,\mathrm{ps}^{-1}$ with one year of data taking. \ATLAS\ and \CMS\
do not have specialized particle identification, but will run at higher luminosities and thus complement \LHCb\
in the analyses of rare decays. Reviews of the current status of B-physics and the \LHCb\ physics programme
can be found in Refs.\ \cite{SchuneLisbon} and \cite{LHCbphysics}.

%\begin{figure}[htbp]
%\begin{center}
%\begin{tabular}{cc}
%\includegraphics[width=7.5cm]{deltams}  &
%\includegraphics[width=6.5cm]{HeavyIonXsec}
%\end{tabular}
%\caption{Left~: Simulation of a $\mathrm{B}_\mathrm{s}$ oscillation measurement with \LHCb.
%               Right~:  Hadron production cross sections as a function of $p_\mathrm{T}$, for various colliders \cite{HIxsec}.}
%\label{BandHIplots}
%\end{center}
%\end{figure}

%%%%%%%%%%%%%%%%%%%%%%%%%%%%%%%%%%%%%%%%

\subsection{The study of heavy ion collisions}
 \label{HI}

The physics of heavy ions and strong phase transitions might enter a real discovery regime
at \LHC, thanks to the higher centre-of-mass energy and a ten-fold increase in the
energy density $\epsilon$ [GeV/fm$^3$] compared to \RHIC\ at Brookhaven. 
A particular feature of \LHC\ is that for the first time high-$p_\mathrm{T}$ objects 
(jets, quarkonia) will be produced in heavy ion collisions,
with huge statistics in a large
variety of processes. Medium effects, such as medium-modified QCD radiation, are
expected to be large and will be studied in detail. The  energy loss of a 
high-$p_\mathrm{T}$ parton, when traversing the dense medium formed in heavy ion collisions,
can be observed as jet quenching. At the same time it is interesting to measure
the low-$p_\mathrm{T}$ tracks in order to get an understanding of the energy flow in these events.
Jets, quarkonia and open flavour production 
are hard probes of the produced dense matter and will hopefully give
unprecedented access to the equilibrium and non-equilibrium QCD dynamics. The \LHC\ 
considerably extends the kinematic range in the small Bjorken-$x$ regime and therefore might
give insight into phenomena such as perturbative saturation. Finally, 
collective phenomena such as radial and elliptic flow are expected to be even stronger
than at \RHIC, where indications have been reported that the formed medium shows strong collective 
behaviour, supporting the picture of an almost ideal liquid \cite{RHICperfectliquid}. 
One of the first (immediate) measurements, already with
about  $10^5$ events, will be the determination of the actual multiplicity per unit rapidity. This rich 
physics programme will be addressed by \ALICE\ and complemented by \CMS\ and \ATLAS.
Qualitatively new experimental tools will be employed, such as very large rapidity coverage, high
granularity and, in the special case of \ALICE, excellent particle identification systems. 
A recent overview and outlook on heavy ion physics is found in Ref.\ \cite{WiedemannHCP}.

%%%%%%%%%%%%%%%%%%%%%%%%%%%%%%%%%%%%%%%%

\section{Physics reach and upgrades}
 \label{later}

Obviously it is very difficult to predict the various physics scenarios after
the first years of \LHC\ running, since this strongly depends on early discoveries (or simple confirmations
of the SM). If one or more Higgs bosons are found, then the next step will be to measure
its (their) parameters (masses, couplings). However, in order to reach the ultimate precision
and sensitivity, an integrated luminosity of up to $300\,\mathrm{fb}^{-1}$ will be needed. 
Most likely the Higgs self-coupling will not be accessible, even with such a large data sample,
and its measurement has to be postponed to a possible Super-\LHC\ (\SLHC).
If there is first evidence for the production of SUSY particles, the following years will be devoted
to the measurement of the sparticle masses, eg.\ by studying cascade decays and the 
related end-points of the lepton spectra. A non-trivial question to be answered will be
the identification of the actually realized SUSY model and the underlying SUSY-breaking
mechanism. Definitely one of the most outstanding successes of \LHC\ would be the
establishment of a clear connection to cosmological questions, eg.\ the discovery 
of a dark matter candidate such as a neutralino. Nevertheless, this has to be complemented
by direct astro-physical observations in order to have a firm understanding of the important
dark matter issue. At \LHC\ it will be difficult or even impossible to observe sleptons with
masses above 350 GeV/$c^2$, to explore the full gaugino mass spectrum, to measure
the spin-parity and couplings of all sparticles and to disentangle squarks of the first
two generations. An international linear collider would definitely offer a complementary approach
to these problems. Finally, if none or only some of the above discoveries are made, all
efforts will be dedicated to the search for other mechanisms of electro-weak symmetry breaking
and/or extra dimensions. In the case no Higgs is found, we will  concentrate on 
the resonant ('easy') or non-resonant ('very difficult') 
scattering of the longitudinal components of two W bosons, since unitarity conservation tells us
that something has to happen there when approaching the TeV energy scale. Ultimately it could
be that we have to accept fine tuning, as for example suggested by "Split SUSY" models \cite{SplitSUSY}.
Whatever nature is going to offer us, we are convinced that the \LHC\ experiments are designed
such that they are also ready for the unexpected. A more detailed discussion of the ultimate
physics reach can be found in Refs.\  \cite{FabiolaHinchcliffe} and \cite{FabiolaLP05}.

Possible upgrade scenarios of the machine and the experiments in view of  \SLHC\ are 
already under study. For the machine, the first step would be to  push its parameters
to the ultimate limits (eg.\ increase the beam energy to 7.45 TeV and the intensity
to $1.7\times10^{11}$ protons/bunch), without any hardware modifications. In a second phase the performance 
would be further improved by modifications in the insertion regions, for 
examples regarding the superconducting triplet quadrupole magnets. The bunch spacing may be
reduced to 25 ns (if the electron cloud effect is under control), the $\beta^*$ lowered to
0.25 m and the beam crossing angle increased to $445\,\mu$rad. Although the peak luminosity
might reach values close to $10^{35} \mathrm{cm}^{-2}\mathrm{s}^{-1}$, future studies will have to
watch closely the luminosity lifetime (efficiency) of the machine, due to an increased energy stored in the
beams and thus increased demands on machine control and protection. A further performance upgrade
in a third phase would only be possible with major hardware modifications.

Also within the experiments the R\&D for a \SLHC\ upgrade has started already now.
The detector upgrades have to take into consideration increased radiation levels and pile-up noise.
A reduced bunch spacing will require hardware modifications for the first-level triggers. The forward regions
will be under particular radiation pressure, thus an improved shielding is under discussion. Generally,
the developments go into the direction of faster and more granular sub-detectors, most notably
new tracking systems. Recent reviews of the requirements and developments are found in 
Refs.\ \cite{GianottiSLHC} and \cite{GreenSLHC}.

%%%%%%%%%%%%%%%%%%%%%%%%%%%%%%%%%%%%%%%%
\newpage

\section{Conclusions}
 \label{conclusion}
 
 A very exciting period for particle physics is lying ahead of us. At the \LHC\ we will
 explore the TeV energy scale for the first time, with a direct discovery potential 
 up to several TeV. We expect to answer many of the current questions 
 regarding the extensions of the SM, and possibly new questions will arise. \CERN\ is
 fully committed to the \LHC\ project. The machine and the detectors will be ready 
 for first beams in summer 2007.
 
%%%%%%%%%%%%%%%%%%%%%%%%%%%%%%%%%%%%%%%%

\section{Acknowledgements}

I would like to thank the conference organizers for the invitation and my many
colleagues from the \LHC\ machine and detector groups for their help in preparing 
this review. I also thank D.\ Treille for his comments on the manuscript.

%%%%%%%%%%%%%%%%%%%%%%%%%%%%%%%%%%%%%%%%

%%%%%%%%%%%%%%%%%%%%%%%%%%%%%%%%%%%%%%%%
%%%%%%%%%%%%%%%%%%%%%%%%%%%%%%%%%%%%%%%%

\end{document}